\documentclass{./styles/svproc}
\pdfoutput=1 
\usepackage{url}

\usepackage{amsmath}
\usepackage{amssymb}
\usepackage{float}
\usepackage{microtype}
\usepackage{graphicx}
\usepackage{orcidlink}
\begin{document}
\mainmatter              
\title{A Portfolio's Common Causal Conditional Risk-Neutral PDE}
\titlerunning{Common Causal Risk-Neutral PDE}  
%
\author{Alejandro Rodriguez Dominguez\inst{1,2} \orcidlink{https://orcid.org/0000-0002-2400-1097}}
\authorrunning{Alejandro Rodriguez Dominguez} 
%
\tocauthor{Alejandro Rodriguez Dominguez}
\institute{Department of Quantitative Research, Miralta Finance Bank S.A., Madrid, Spain, \email{arodriguez@miraltabank.com}
\and
Department of Computer Science, University of Reading, 
Whiteknights House, Reading, United Kingdom}

\maketitle              

\begin{abstract}
Portfolio's optimal drivers for diversification are common causes of the constituents' correlations. A closed-form formula for the conditional probability of the portfolio given its optimal common drivers is presented, with each pair constituent-common driver joint distribution modelled by Gaussian copulas. A conditional risk-neutral PDE is obtained for this conditional probability as a system of copulas’ PDEs, allowing for dynamical risk management of a portfolio as shown in the experiments. Implied conditional portfolio volatilities and implied weights are new risk metrics that can be dynamically monitored from the PDEs or obtained from their solution.

\end{abstract}
\section{Common Causal Conditional Risk-Neutral PDE}
It has been proven that optimal portfolio drivers' for diversification must be the common causes of portfolio constituents' correlations \cite{RODRIGUEZDOMINGUEZ2023100447}. Reichenbach’s Common Cause Principle (RCCP) provides a set of independent conditions that variables need to satisfy to be a common cause of a probabilistic correlation \cite{Reichenbach1956-REITDO-2}. Assume that $(\Omega, \mathcal{F}, P)$ is a standard probability space representing a financial market. A portfolio consists of n financial assets $a_t=[a_{1t},\dots,a_{nt}]$ at time $t$ with respective weights $\boldsymbol{w}$. The optimal common causal drivers for the portfolio are selected based on the Commonality Principle \cite{RODRIGUEZDOMINGUEZ2023100447}. Specifically, the subset of $m$ optimal common causal drivers for a portfolio are selected so that the Reichembach's Common Cause Principle independent conditions have the highest probability \cite{Reichenbach1956-REITDO-2}, from a set of drivers' candidates $M$ belonging to $\Omega$ with $M>>m$. The subset of optimal portfolio common drivers $\boldsymbol{D}=\boldsymbol{D_{t-1}}=[D_{t-1,1},\dots,D_{t-1,m}]$ assuming for simplicity a lag of 1, satisfying the RCCP conditions, make the portfolio constituents conditional on $\boldsymbol{D}$, independent:

\begin{equation}
    P(p_t|\boldsymbol{{D}_{t-1}})=P\left(\sum_{i=1}^{n}{w_i a_{it}}\middle| \boldsymbol{{D}_{t-1}}\right)=\sum_{i=1}^{n}\sum_{j=1}^{m}{w_i{{D}_{t-1,j}P\left(a_{it}\middle|{D}_{t-1,j}\right)}}\ \ \forall t
    \label{eq1}
\end{equation}
Jeffrey conditionalization, $P\left(a_{it}|{D}_{t-1,1}\equiv {D}_{t-1,1},\dots,{D}_{t-1,m}\equiv {D}_{t-1,m}\right)=\\
=\sum_{j=1}^{m}{{D}_{t-1,j}P(a_{it}|{D}_{t-1,j})}$, $\forall t, \forall i=1,\dots n$, is applied in the final step of (\ref{eq1}). The joint probabilities of each constituent, with respect to each common driver, follow bivariate distributions that can be modelled with copulas. In the case of a Gaussian copula, with $a_i=a_{it}$, $D_j=D_{t-1,j}$, for any particular $t$, the density is:
\begin{align}
    &C\left(a_i,D_j\right)=\nonumber\\
    &=\frac{1}{\sqrt{1-\rho_{ij}^2}}exp\left(-\frac{\rho_{ij}^2\left({\mathrm{\Phi}^{-1}\left(a_i\right)}^2+{\mathrm{\Phi}^{-1}\left(D_j\right)}^2\right)-2\rho_{ij}\mathrm{\Phi}^{-1}\left(a_i\right)\mathrm{\Phi}^{-1}\left(D_j\right)}{2\left(1-\rho_{ij}^2\right)}\right)
    \label{eq2}
\end{align}
The conditional probability is given by:
\begin{equation}
    P(a_i\le a_i\ |\ D_j=D_j)=\left.\frac{\partial}{\partial D_j}C(a_i,D_j)\right|_{\left(F_{a_i}(a_i),F_{D_j}(D_j)\right)}
    \label{eqpart}
\end{equation}
Applying (\ref{eq2}) and (\ref{eqpart}) to (\ref{eq1}), $\forall t$:
\begin{equation}
    P(p_t|\boldsymbol{{D}_{t-1}})=\sum_{i=1}^{n}\sum_{j=1}^{m}{w_i{D}_{t-1,j}}\frac{\partial}{\partial {D}_{t-1,j}}C(F_{a_{it}}(a_{it}),F_{{D}_{t-1,j}}({D}_{t-1,j}))
    \label{eq3}
\end{equation}
Using (\ref{eq3}), and by computing the partial derivatives of C from (\ref{eq2}):
\begin{align}
&P(p|\boldsymbol{D})=\nonumber\\
&=\sum_{i=1}^{n}\sum_{j=1}^{m}\frac{-{w_i{D}_j} \exp\left(\frac{-\ {\rho_{a_i{D}_j}}^2\left({x_1}^2+{x_2}^2\right)-2\rho_{a_i{D}_j}x_1x_2}{2\left(1-{\rho_{a_i{D}_j}}^2\right)}\right)2{\rho_{a_i{D}_j}}^2\frac{x_2}{\mathrm{\Phi}^\prime\left(x_2\right)}-2\rho_{a_i{D}_j}\frac{x_1}{\mathrm{\Phi}^\prime\left(x_2\right)}}{2\left(1-{\rho_{a_i{D}_j}}^2\right)^{3/2}}
\label{eq7}
\end{align}
with $p=p_t$, $\boldsymbol{D}=\boldsymbol{{D}_{t-1}}$, $a_i=a_{it}$, ${D}_j={D}_{t-1,j}$, $x_1=\mathrm{\Phi}^{-1}(a_i)$ and $x_2=\mathrm{\Phi}^{-1}({D}_j)$. In matrix form (\ref{eq7}) is given by:
\begin{equation}
    P(p|\boldsymbol{D})=-\boldsymbol{w^T}\boldsymbol{\Pi} \boldsymbol{D}
    \label{eq8}
\end{equation}
with $\boldsymbol{w}=\left[\begin{matrix}w_1\\\vdots\\w_n\\\end{matrix}\right]$, $\boldsymbol{\Pi}_{(nxm)}=\left[\begin{matrix}\frac{\partial C(a_1,{D}_1)}{\partial{D}_1}&\ldots&\frac{\partial C(a_1,{D}_m)}{\partial{D}_m}\\\vdots&\ddots&\vdots\\\frac{\partial C(a_n,{D}_1)}{\partial{D}_1}&\ldots&\frac{\partial C(a_n,{D}_m)}{\partial{D}_m}\\\end{matrix}\right]$ and $\boldsymbol{D}_{(mxn)}=\left[\begin{matrix}{D}_1&\ldots&{D}_1\\\vdots&\ddots&\vdots\\{D}_m&\ldots&{D}_m\\\end{matrix}\right]$. Deriving (\ref{eq8}) with respect to $t$, the following PDE is obtained:
\begin{equation}
    \frac{\partial\ P(p|\boldsymbol{D})}{\partial t}dt=-\boldsymbol{w^T}\left[\left(\frac{\partial\boldsymbol{\Pi}}{\partial \boldsymbol{a}}\frac{\partial \boldsymbol{a}}{\partial t}+\frac{\partial\boldsymbol{\Pi}}{\partial \boldsymbol{D}}\frac{\partial \boldsymbol{D}}{\partial t}+\frac{\partial\boldsymbol{\Pi}}{\partial\boldsymbol{\rho}}\frac{\partial\boldsymbol{\rho}}{\partial t}\right)\boldsymbol{D}+\boldsymbol{\Pi}\frac{\partial \boldsymbol{D}}{\partial t}\right]dt
\end{equation}
which can be expressed in terms of $P(p|\boldsymbol{D})$ by using (\ref{eq8}) as:
\begin{equation}
    \left[\frac{\partial\ P(p|\boldsymbol{D})}{\partial t}+\frac{\partial\ P(p|\boldsymbol{D})}{\partial p}\frac{\partial p}{\partial t}+\frac{\partial\ P(p|\boldsymbol{D})}{\partial \boldsymbol{D}}\frac{\partial \boldsymbol{D}}{\partial t}+\left(\frac{\partial\ P(p|\boldsymbol{D})}{\partial \boldsymbol{\rho}}-\boldsymbol{w^T}\boldsymbol{\Pi}\frac{\partial \boldsymbol{D}}{\partial \boldsymbol{\rho}}\right)\frac{\partial \boldsymbol{\rho}}{\partial t}\right]dt=0
    \label{PDE2}
\end{equation}
and the partial derivatives can also be obtained from (\ref{eq8}) as $\frac{\partial\ P(p|\boldsymbol{D})}{\partial p}=\boldsymbol{w^T}\left(\frac{\partial\boldsymbol{\Pi}}{\partial \boldsymbol{a}}\boldsymbol{D}\right)$, $\frac{\partial\ P(p|\boldsymbol{D})}{\partial \boldsymbol{D}}=\boldsymbol{w^T}\left(\frac{\partial\boldsymbol{\Pi}}{\partial \boldsymbol{D}}\boldsymbol{D}+\boldsymbol{\Pi}\right)$ and $\frac{\partial\ P(p|\boldsymbol{D})}{\partial \boldsymbol{\rho}}=\boldsymbol{w^T}\left(\frac{\partial\boldsymbol{\Pi}}{\partial\boldsymbol{\rho}}\boldsymbol{D}+\boldsymbol{\Pi}\frac{\partial \boldsymbol{D}}{\partial\boldsymbol{\rho}}\right)$. Portfolio dynamics $\frac{\partial p}{\partial t}$ are given by the Ito process $dp_t=\mu_pdt+\sigma_pdW_{p_t}$ with $\sigma_p$ the volatility. The common drivers' dynamics $\frac{\partial \boldsymbol{D}}{\partial t}$ follow the m-dimensional Ito process $\boldsymbol{dD_t}=\boldsymbol{\mu_D}dt+\boldsymbol{\sigma_D}d\boldsymbol{W_{D_t}}$, with $\boldsymbol{\sigma_D}$ the square-root of the variance-covariance matrix of the common drivers, and $\boldsymbol{W_{D_t}}$ a m-dimensional correlated Brownian motion. By applying Ito's lemma to $P(p|\boldsymbol{D})$:
\begin{align}
  &dP(p|\boldsymbol{D})=\left[\frac{\partial P(p|\boldsymbol{D})}{\partial t}+\frac{1}{2}\left(\frac{\partial^2P(p|\boldsymbol{D})}{{\partial p}^2}\sigma_p^2+\frac{\partial^2P(p|\boldsymbol{D})}{{\partial{\rm \boldsymbol{D}}}^2}\boldsymbol{\sigma_{D}}^2\right)+\right.\nonumber\\
  &\left.+\frac{\partial^2P(p|\boldsymbol{D})}{\partial p\partial{\rm \boldsymbol{D}}}\sigma_{p}\boldsymbol{\sigma_{D}}\right]dt+\frac{\partial P(p|\boldsymbol{D})}{\partial p}{dp_t}+\frac{\partial P(p|\boldsymbol{D})}{\partial{\rm \boldsymbol{D}}}d\boldsymbol{{\rm D}_t}
  \label{ito}
\end{align}
By substituting the PDE given in (\ref{PDE2}) into the Ito's lemma derivation of $P(p|\boldsymbol{D})$ in (\ref{ito}):
\begin{align}
    &dP(p|\boldsymbol{D})=\left[\frac{1}{2}\left(\frac{\partial^2P(p|\boldsymbol{D})}{{\partial p}^2}\sigma_{p}^2+\frac{\partial^2P(p|\boldsymbol{D})}{{\partial{\rm \boldsymbol{D}}}^2}\boldsymbol{\sigma_{D}}^2\right)+\frac{\partial^2P(p|\boldsymbol{D})}{\partial p\partial{\rm \boldsymbol{D}}}\sigma_{p}\boldsymbol{\sigma_{D}}\right]dt-\nonumber\\
    &-\left(\frac{P(p|\boldsymbol{D})}{\boldsymbol{D}}\frac{\partial\ \boldsymbol{D} }{\partial \boldsymbol{\rho}}+\frac{\partial\ P(p|\boldsymbol{D})}{\partial \boldsymbol{\rho}}\right)d\boldsymbol{\rho_t}
    \label{Eq13}
\end{align}
with $\boldsymbol{\rho_t}$ a $(n*m)$-dimensional vector containing all the correlations between portfolio constituents and the common drivers (not a correlation matrix). These correlations' dynamics must follow Ito processes of any suitable kind, which is irrelevant for this work as the expressions become simplified, but including a $(n*m)$-dimensional Brownian motion. For simplicity, assuming it follows a $(n*m)$-dimensional geometric Brownian motion $\frac{\partial \boldsymbol{\rho}}{\partial t}=d\boldsymbol{\rho_t}=\boldsymbol{\mu_{\rho}}\boldsymbol{\rho_t} dt+\boldsymbol{\sigma_{\rho}}\boldsymbol{\rho_t} d\boldsymbol{W_{\rho}^t}$, substituting in (\ref{Eq13}):
\begin{align}
    &dP(p|\boldsymbol{D})=\left[\frac{1}{2}\left(\frac{\partial^2P(p|\boldsymbol{D})}{{\partial p}^2}\sigma_{p}^2+\frac{\partial^2P(p|\boldsymbol{D})}{{\partial\boldsymbol{{\rm D}}}^2}\boldsymbol{\sigma_{D}}^2\right)+\frac{\partial^2P(p|\boldsymbol{D})}{\partial p\partial{\rm \boldsymbol{D}}}\sigma_{p}\boldsymbol{\sigma_{D}}-\right.\nonumber\\
    &\left.-\left(\frac{P(p|\boldsymbol{D})}{\boldsymbol{D}}\frac{\partial\ \boldsymbol{D}}{\partial \boldsymbol{\rho}}+\frac{\partial\ P(p|\boldsymbol{D})}{\partial \boldsymbol{\rho}}\right)\boldsymbol{\mu_{\rho}}\boldsymbol{\rho_t}\right]dt-\left(\frac{P(p|\boldsymbol{D})}{\boldsymbol{D}}\frac{\partial\ \boldsymbol{D}}{\partial \boldsymbol{\rho}}+\frac{\partial\ P(p|\boldsymbol{D})}{\partial \boldsymbol{\rho}}\right)\boldsymbol{\sigma_{\rho}}\boldsymbol{\rho_t} d\boldsymbol{W_{\rho}^t}
    \label{eq14}
\end{align}
The common causal conditional risk-neutral PDE is given by equating the drift part in (\ref{eq14}) to the drift of the common causal drivers' processes' times the conditional probability of the portfolio given these set of common causal drivers. This is due to the fact that, when portfolio constituents are projected in a common causal conditional probability space, they become independent and the portfolio risk is diversified, therefore the common causal conditional probability space becomes conditionally risk-neutral. The return of the portfolio conditional on the common drivers is proportional to the common drivers' drift term plus an error term $\Delta$ accounting for portfolio's common causal drivers' error selection:
\begin{align}
\left[\frac{1}{2}\left(\frac{\partial^2P(p|\boldsymbol{D})}{{\partial p}^2}\sigma_p^2+\frac{\partial^2P(p|\boldsymbol{D})}{{\partial\boldsymbol{{\rm D}}}^2}\boldsymbol{\sigma_{D}}^2\right)+\frac{\partial^2P(p|\boldsymbol{D})}{\partial p\partial{\rm \boldsymbol{D}}}\sigma_p\boldsymbol{\sigma_{D}}\right]dt=\left[\boldsymbol{\mu_{D}} P(p|\boldsymbol{D})+\boldsymbol{\Delta}\right]dt
\label{eq15}
\end{align}
and there is one condition, given by the following PDE, such that the PDE in (\ref{eq14}) does not have a Brownian motion component, making it conditional risk-neutral:
\begin{align}
\frac{P(p|\boldsymbol{D})}{\boldsymbol{D}}\frac{\partial \boldsymbol{D}}{\partial \boldsymbol{\rho}}= - \frac{\partial P(p|\boldsymbol{D})}{\partial \boldsymbol{\rho}}
\label{eq16}
\end{align}
Parameter $\boldsymbol{\Delta}$ in (\ref{eq15}) measures the deviations from conditional risk neutrality due to unknown confounders or not having selected all optimal common causal drivers for the portfolio. We assume now for simplicity it is zero. It can work as a loss function for a learning/calibration problem in which the optimal set of common drivers are found when $\boldsymbol{\Delta}$ is minimum. Additionally, it can be used as a signal that the common drivers' set needs to be revised. If the partial derivatives computed from (\ref{eq8}) are substituted in (\ref{eq16}):
\begin{align}
\frac{-\boldsymbol{w^T}\boldsymbol{\Pi} \boldsymbol{D}}{\boldsymbol{D}}\frac{\partial \boldsymbol{D}}{\partial \boldsymbol{\rho}}=-\boldsymbol{w^T}\left(\frac{\partial\boldsymbol{\Pi}}{\partial\boldsymbol{\rho}}\boldsymbol{D}+\boldsymbol{\Pi}\frac{\partial \boldsymbol{D}}{\partial\boldsymbol{\rho}}\right)
\end{align}
which simplifies to:
\begin{align}
\boldsymbol{\Pi}\frac{\partial \boldsymbol{D}}{\partial \boldsymbol{\rho}}=\left(\frac{\partial\boldsymbol{\Pi}}{\partial\boldsymbol{\rho}}\boldsymbol{D}+\boldsymbol{\Pi}\frac{\partial \boldsymbol{D}}{\partial\boldsymbol{\rho}}\right)
\end{align}
This gives another condition for the PDE, $\frac{\partial \boldsymbol{\Pi}}{\partial \boldsymbol{\rho}}\boldsymbol{D}=0$, which can be derived analytically from $\boldsymbol{\Pi}$ in (\ref{eq8}), and consists of a system of n equations. From now on, the common causal conditional risk-neutral PDE is expressed in terms of the vector of portfolio constituents' returns, $\boldsymbol{a}$, instead of the portfolio's return $p$ in (\ref{eq15}). The partial derivatives of $P(p|\boldsymbol{D})$ in (\ref{eq15}) are computed using the expression in (\ref{eq8}):
\begin{align}
  \frac{\partial^2\ P(p|\boldsymbol{D})}{\partial p^2}=\boldsymbol{w^T}\left(\frac{\partial^2\boldsymbol{\Pi}}{\partial \boldsymbol{a}^2}\boldsymbol{D}+\frac{\partial\boldsymbol{\Pi}}{\partial \boldsymbol{a}}\frac{\partial\boldsymbol{D}}{\partial \boldsymbol{a}}+\frac{\partial\boldsymbol{\Pi}}{\partial \boldsymbol{a}}\frac{\partial \boldsymbol{D}}{\partial \boldsymbol{a}}+\boldsymbol{\Pi}\frac{\partial^2 \boldsymbol{D}}{\partial \boldsymbol{a}^2}\right)=\boldsymbol{w^T}\left(\frac{\partial^2\boldsymbol{\Pi}}{\partial \boldsymbol{a}^2}\boldsymbol{D}\right)  
  \label{Eq166}
\end{align}
\begin{align}
 \frac{\partial^2\ P(p|\boldsymbol{D})}{\partial \boldsymbol{D}^2}=\boldsymbol{w^T}\left(\frac{\partial^2\boldsymbol{\Pi}}{\partial \boldsymbol{D}^2}\boldsymbol{D}+2\frac{\partial\boldsymbol{\Pi}}{\partial \boldsymbol{D}}\right); \frac{\partial^2\ P(p|\boldsymbol{D})}{\partial \boldsymbol{\rho}^2}=\boldsymbol{w^T}\left(\frac{\partial\boldsymbol{\Pi}}{\partial\boldsymbol{\rho}}\frac{\partial \boldsymbol{D}}{\partial\boldsymbol{\rho}}+\boldsymbol{\Pi}\frac{\partial^2 \boldsymbol{D}}{\partial\boldsymbol{\rho}^2}\right)
 \label{Eq176}
 \end{align}
 \begin{align}
 \frac{\partial^2P(p|\boldsymbol{D})}{\partial p\partial{\rm \boldsymbol{D}}}=\boldsymbol{w^T}\left(\frac{\partial\boldsymbol{\Pi}}{\partial \boldsymbol{a}}+\frac{\partial\boldsymbol{\Pi}}{\partial \boldsymbol{a}\partial{\rm \boldsymbol{D}}}\boldsymbol{D}\right)
 \label{Eq186}
\end{align}
substituting (\ref{Eq166}-\ref{Eq186}) in (\ref{eq15}), and assuming $\boldsymbol{\Delta}=0$:
\begin{align}
&\boldsymbol{w^T}\left[\frac{1}{2}\left(\left(\frac{\partial^2\boldsymbol{\Pi}}{\partial \boldsymbol{a}^2}\boldsymbol{D}\right)\boldsymbol{\sigma_p}^2+\left(\frac{\partial^2\boldsymbol{\Pi}}{\partial \boldsymbol{D}^2}\boldsymbol{D}+2\frac{\partial\boldsymbol{\Pi}}{\partial \boldsymbol{D}}\right)\boldsymbol{\sigma_{D}}^2\right)+\left(\frac{\partial\boldsymbol{\Pi}}{\partial \boldsymbol{a}}+\frac{\partial\boldsymbol{\Pi}}{\partial \boldsymbol{a}\partial{\rm \boldsymbol{D}}}\boldsymbol{D}\right)\boldsymbol{\sigma_p}\boldsymbol{\sigma_{D}}\right]=\nonumber\\
&=-\boldsymbol{\mu_{D}}\boldsymbol{w^T}\boldsymbol{\Pi} \boldsymbol{D}
\label{eq24}
\end{align}
with the portfolio variance ${\boldsymbol{\sigma_p}}^T\boldsymbol{\sigma_p}=\boldsymbol{w}^T\boldsymbol{\Sigma_p} \boldsymbol{w}=\boldsymbol{\sigma_p}^2$ now expressed as a diagonal variance-covariance matrix of independent portfolio constituents, and having factorized (\ref{eq24}) by $\boldsymbol{w}^T$, getting, $\boldsymbol{\Sigma_p} \boldsymbol{w}=\left[\begin{matrix}\sigma_1^2&\ldots&0\\\vdots&\ddots&\vdots\\0&\ldots&\sigma_n^2\\\end{matrix}\right]\boldsymbol{w}$. A PDE that does not depend on the weights of the portfolio, except from the term $\boldsymbol{\Sigma_p} \boldsymbol{w}$, is obtained:
\begin{align}
&\frac{1}{2}\left(\left(\frac{\partial^2\boldsymbol{\Pi}}{\partial \boldsymbol{a}^2}\boldsymbol{D}\right)\boldsymbol{\Sigma_p} \boldsymbol{w}+\left(\frac{\partial^2\boldsymbol{\Pi}}{\partial \boldsymbol{D}^2}\boldsymbol{D}+2\frac{\partial\boldsymbol{\Pi}}{\partial \boldsymbol{D}}\right)\boldsymbol{\sigma_{D}}^2\right)+\left(\frac{\partial\boldsymbol{\Pi}}{\partial \boldsymbol{a}}+\frac{\partial\boldsymbol{\Pi}}{\partial \boldsymbol{a}\partial{\rm \boldsymbol{D}}}\boldsymbol{D}\right)\boldsymbol{\Sigma_p} \boldsymbol{w}\boldsymbol{\sigma_{D}}+\nonumber\\
&+\boldsymbol{\mu_{D}}^T\boldsymbol{D}\boldsymbol{\Pi}=0
\label{eq25}
\end{align}
with $\boldsymbol{\sigma_{D}}^2=\boldsymbol{\Sigma_{D}}=Cov(D_{t-1}^k,D_{t-1}^q)=e^{(\mu_{Dk}+\mu_{Dq})(t-1)}(e^{\rho_{Dk,Dq}\sigma_{Dk}\sigma_{Dq}t}-1)$, $\boldsymbol{\Sigma_p}=Cov(a_t^v,a_t^z)=e^{(\mu_{av}+\mu_{az})t}(e^{\sigma_{av}\sigma_{az}t}-1)$, $\forall k,q = 1,\dots,m; v,z = 1,\dots,n$, and $Cov(a_t^v,a_t^z)=0$ if $v\neq z$. Expressions are given for the partial derivatives in (\ref{eq25}), which can be computed with Kronecker products and substituted back in (\ref{eq25}). One case is shown due to space constraints:
\begin{align}
    &\boldsymbol{D}^T\frac{\partial^2\boldsymbol{\Pi}}{\partial \boldsymbol{a}^2}\boldsymbol{\Sigma_p}\boldsymbol{w} =\nonumber\\
    &\boldsymbol{D}^T_{(nxm)}\left(\left(\frac{\partial}{\partial a_1},\dots,\frac{\partial}{\partial a_n}\right)\oplus\left(\frac{\partial}{\partial a_1},\dots,\frac{\partial}{\partial a_n}\right)\oplus\boldsymbol{\Pi}_{(nxm)}\right)_{(nx(n^2m))}\boldsymbol{\Sigma_p}\boldsymbol{w}_{(nxnx1)}=\nonumber\\
    &=\boldsymbol{D}^T_{(nxm)}\left[\begin{matrix}\frac{\partial^3 C(a_1,{\rm D}_1)}{\partial{\rm D}_1\partial a_1^2}&\ldots&0\\\vdots&\ddots&\vdots\\\frac{\partial^3 C(a_1,{\rm D}_m)}{\partial{\rm D}_m\partial a_1^2}&\ldots&0\\\end{matrix}\begin{matrix}\\\end{matrix}\begin{matrix}\ldots&\frac{\partial^3 C(a_n,{\rm D}_1)}{\partial{\rm D}_1\partial a_n^2}\\\ddots&\vdots&\\\ldots&\frac{\partial^3 C(a_n,{\rm D}_m)}{\partial{\rm D}_m\partial a_n^2}\\\end{matrix}\right]_{(mx(mn^2)}\boldsymbol{\Sigma_p}\boldsymbol{w}_{(n^2x1)}
\end{align}
\begin{figure}[ht] 
\setlength\abovecaptionskip{0\baselineskip}
  
  \begin{minipage}[b]{0.5\linewidth}  \includegraphics[width=1\linewidth]{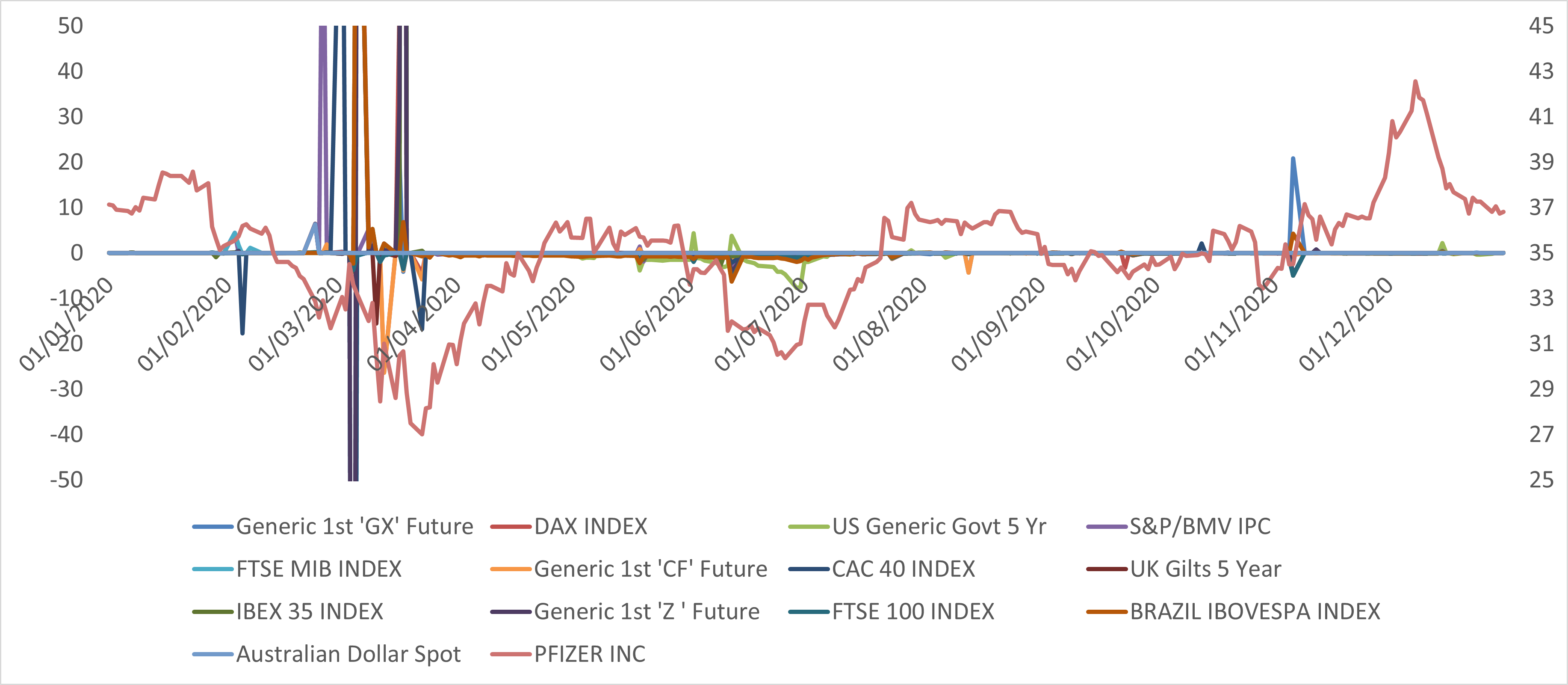} 
  \caption{Copulas' PDEs values for Phyzer - 2020}
  \label{fig1} 
  \end{minipage}
  \begin{minipage}[b]{0.5\linewidth}
    \centering
    \includegraphics[width=1\linewidth]{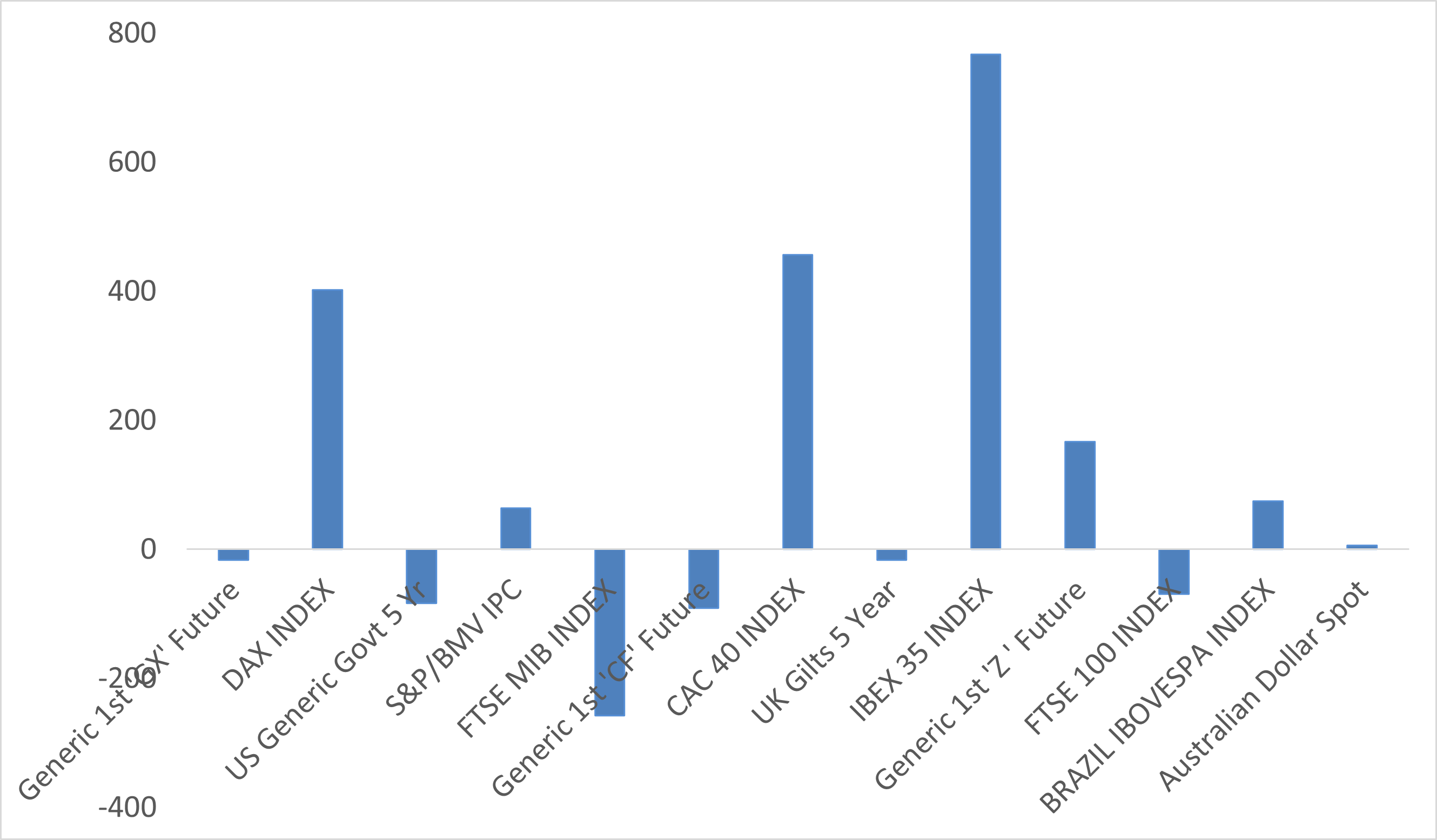} 
    \caption{Sum of values for 2020 (Phyzer)} 
  \label{fig2} 
  \end{minipage} 
  \begin{minipage}[b]{0.5\linewidth}
    \includegraphics[width=1\linewidth]{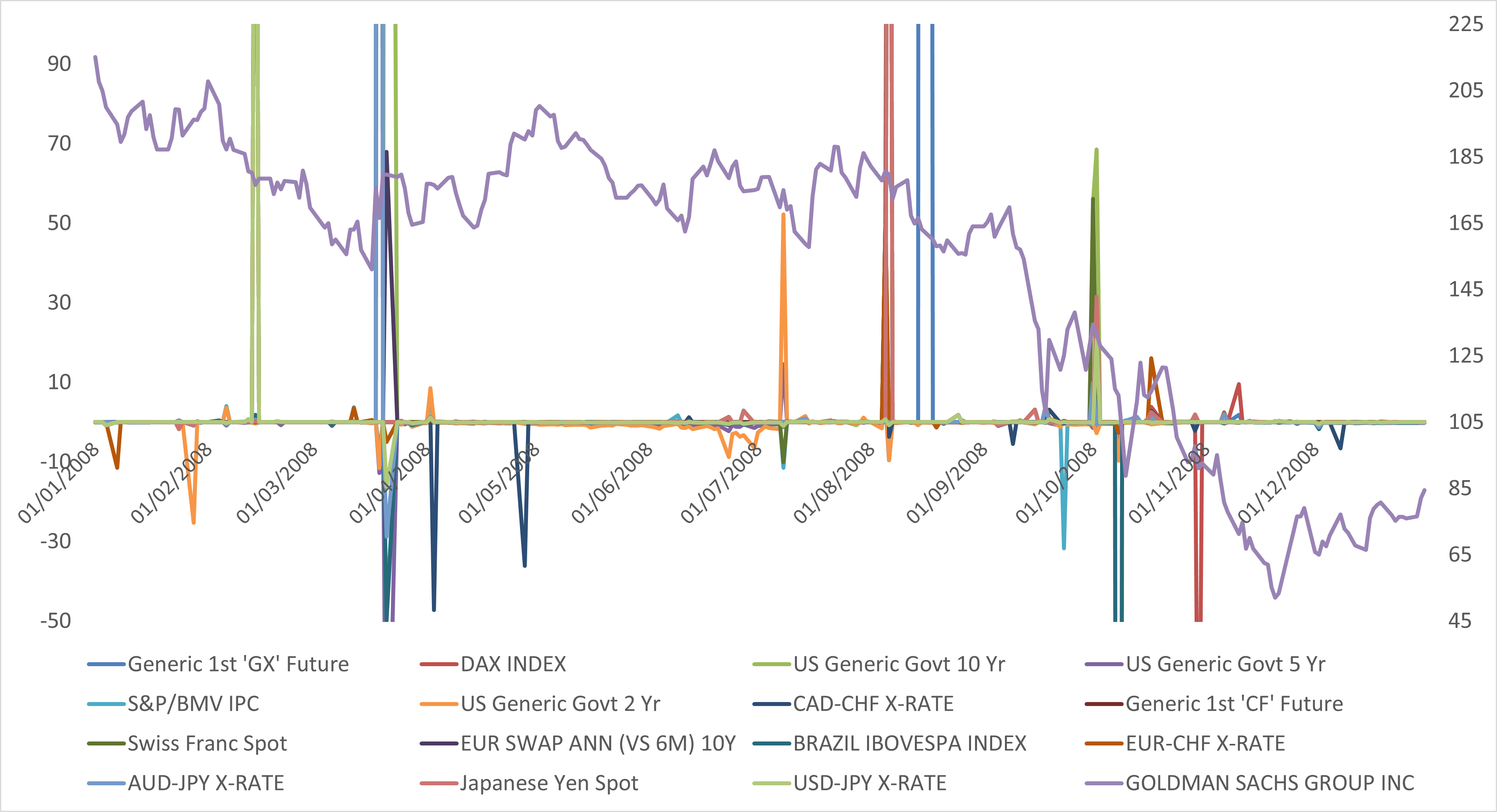} 
    \caption{Copulas' PDEs values for Goldman \\
    Sachs (GS) - 2008} 
    \label{fig3} 
  \end{minipage}
  \begin{minipage}[b]{0.5\linewidth}
    \includegraphics[width=1\linewidth]{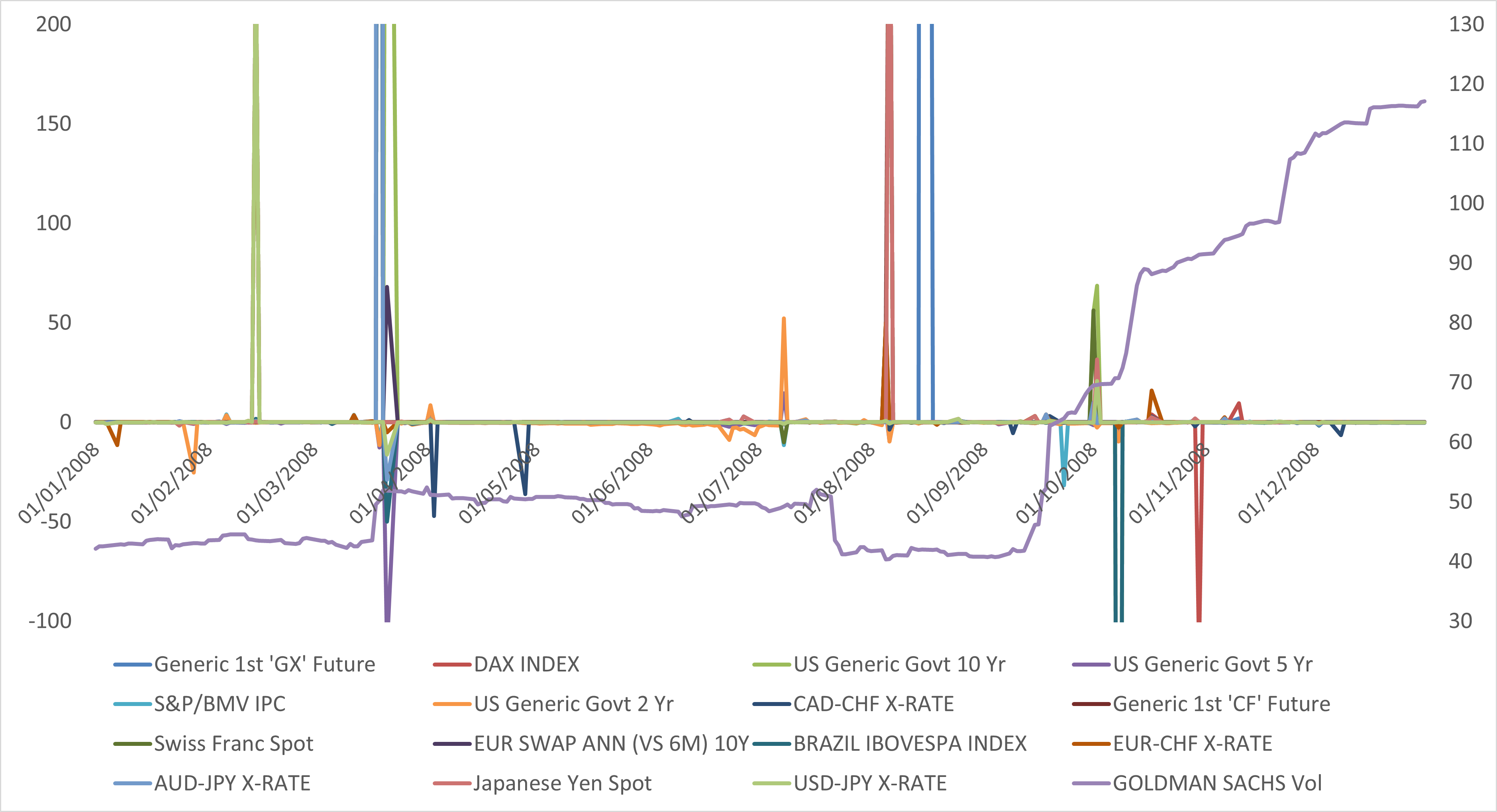} 
    \caption{GS 2008 Volatility} 
    \label{fig4} 
  \end{minipage} 
\end{figure}

\paragraph{Notes and Comments.}
A system of copulas’ PDEs for each pair of constituent/ common driver is obtained. Each PDE can be used for online risk management at constituent level, as seen in Figures \ref{fig1}-\ref{fig4}. Copulas’ PDEs are computed from data on a rolling window basis. Deviations from theoretical values are shown in these figures for two constituents and portfolio’s common drivers. Large deviations are due to important events such as earning reports (From 18/03/2008 earnings reports, GS shares rise more than 30$\%$), idiosyncratic news (From 28/10/2008 the Volkswagen scandal caused GS shares to plunge), and systematic market moves (30/10/2008, Dax rose 2,44$\%$ in one day), all from Figure \ref{fig1}. Figures \ref{fig1} and \ref{fig3} show that deviations remain close to zero except for punctual events, so that the conditional probabilities are almost always described accurately by the PDEs. All copulas' PDEs deviations add to the deviations from conditional risk neutrality in the PDE system expressed as $\boldsymbol{\Delta}$ in (\ref{eq15}). If deviations are added up over a period as in Figure \ref{fig3}, risk-hedging information is obtained and can be included in sensitivity-based methods for portfolio optimization with respect to common causal drivers \cite{RODRIGUEZDOMINGUEZ2023100447}, as an extra component of the sensitivities for tail-risk management (large deviations). Experiments use equal-weight portfolios, but can be selected to minimize the deviations, left as future work. Large deviations are related to increasing volatility (Figure \ref{fig4}). Implied market hedges are obtained after solving the weights in the PDE system. With the system PDEs solution, which is left for future work, implied weights based on the conditional risk-neutral common causal measure could be obtained for longer horizons.

\section*{Acknowledgement}
Thanks to Hugo Valle Valcarcel for helping in the numerical computation of the derivatives and to Miralta Finance Bank S.A. for providing datasets.

%
%

\end{document}